\documentclass[cameraready]{Interspeech}

\title{
Multi-Phonation Graph Learning with Self-Supervised Speech Embeddings for ALS Detection and Progression Prediction}

\author[affiliation={1}, orcid=0000-0003-4031-9537, correspondingauthor]{Behrad}{TaghiBeyglou}
\author[affiliation={1,2}, orcid=0009-0001-7860-0992]{Fatemeh}{Bagheri}
\author[affiliation={1,2}]{Ervin}{Sejdić}

\address{
    $^1$ Department of Electrical and Computer Engineering, University of Toronto, Canada \\
    $^2$ North York General Hospital, Canada
}

\email{\{behrad.taghibeyglou,fatemeh.bagheri\}@mail.utoronto.ca, ervin.sejdic@utoronto.ca}


\usepackage{comment}

\usepackage[utf8]{inputenc}
\usepackage[T1]{fontenc}
\usepackage{amsmath,amssymb}
\usepackage{graphicx}
\usepackage{booktabs}
\usepackage{multirow}
\usepackage{array}
\usepackage{xcolor}
\usepackage{url}
\usepackage{microtype}
\usepackage{hyperref}
\usepackage{cleveref}

\hypersetup{
  colorlinks=true,
  linkcolor=blue,
  citecolor=blue,
  urlcolor=blue
}

\usepackage[includehead,includefoot]{geometry}
\usepackage{pdfpages}

\usepackage{graphicx}
\usepackage{subfigure}
\usepackage{subcaption}

\usepackage{multicol}
\usepackage{multirow}

\usepackage[table]{xcolor}

\newcommand{\bestcell}[1]{\cellcolor{gray!18}\textbf{#1}}
\newcommand{\bestoverall}[1]{\cellcolor{orange!25}\textbf{#1}}

\newcolumntype{C}[1]{>{\centering\arraybackslash}p{#1}}

\usepackage[table]{xcolor} 
\definecolor{lightgray}{gray}{0.9}

\begin{document}

\maketitle

\begin{abstract}
Amyotrophic lateral sclerosis (ALS) progressively impairs speech motor control, making acoustic analysis a promising biomarker for severity and progression estimation. We propose a subject-level graph framework that aggregates multiple phonation recordings into a unique k-nearest-neighbor graph built from pretrained SSL embeddings of 2s segments. We compare four SSL front-ends (wav2vec 2.0, HuBERT, data2vec-audio, and UniSpeech-SAT) and five graph neural networks (GCN, residual GCN, GAT, GraphSAGE, and GIN) on the SAND dataset tasks (339 participants: 205 ALS, 134 control): 5-class dysarthria severity and 4-class ALSFRS-R progression prediction. On the official validation set, the best configuration (HuBERT+GIN) achieves macro-F$_1$ of 0.73 for Task 1 and 0.69 for Task 2, outperforming SAND validation baselines (0.61 and 0.58). These results highlight the potential of combining GNNs with pretrained cross-lingual speech representations for low-resource ALS detection and progression monitoring.
\end{abstract}


\section{Introduction}
\label{sec:intro}

Amyotrophic lateral sclerosis (ALS) is a fatal neurodegenerative disease that
progressively impairs motor function, frequently affecting speech early in the
disease course~\cite{kiernan2011als}. Dysarthria-related changes in sustained
vowel phonation and diadochokinesis (DDK) are clinically informative and can be
captured non-invasively with brief voice tasks, motivating speech as a scalable
digital biomarker for monitoring and assessment~\cite{verde2016vox4health,dubbioso2024voice}.
However, robust automated staging and progression prediction remain challenging
due to limited labeled clinical data, high inter-speaker variability, and the
fact that clinically relevant cues may be distributed across multiple elicitation
tasks and time segments.

Prior ALS speech-assessment pipelines have largely followed two directions.
First, many systems rely on handcrafted acoustic descriptors capturing
phonation, articulation, and time-frequency structure (e.g., jitter/shimmer,
harmonic-to-noise-ratio, pitch and spectral features) paired with classical machine learning
classifiers to detect bulbar involvement or differentiate clinical
subgroups~\cite{tena2022bulbar}. Second, deep models trained directly on audio
or spectro-temporal inputs have improved predictive accuracy and enabled
post-hoc interpretability, but they remain data-reliant, which makes them less generalizable from one language or accent to the other, and accordingly they typically require
careful task- and dataset-specific training to generalize across speakers,
recording conditions, and elicitation protocols~\cite{merler2025npj}. More
broadly across dysarthria severity modeling, comparative studies highlight that
performance is sensitive to feature design and architecture choice, reinforcing
the limitation of relying on limited labeled data and narrow acoustic feature
sets~\cite{joshy2022tnsre}.


Recent self-supervised learning (SSL) speech models provide transferable
representations that reduce the need for task-specific labels and have shown
strong generalization across many downstream speech problems~\cite{mohamed2022sslreview,yang2021superb}.
In parallel, graph neural networks (GNNs) offer a principled mechanism for
aggregating information over sets of related items through message passing.
Motivated by these advances, we view a subject's multiple short speech segments
as a structured set and explicitly model their relationships in embedding
space. Previous studies have looked into the potential of graph-based speech modeling for pathological speech \cite{cai2022gnn_pathological} and Parkinson's disease \cite{sheikh2025graph}; however, their potentials have not been explored in field of ALS dysarthria severity estimation. 

In this work, we propose a subject-level graph learning pipeline that (1)
extracts frozen SSL embeddings from short (2\,s) segments using established SSL
front-ends~\cite{baevski2020wav2vec2,hsu2021hubert,baevski2022data2vec,chen2022unispeechsat},
(2) builds a per-subject $k$-nearest-neighbor (kNN) graph using cosine similarity, and
(3) performs graph classification with standard GNN backbones. Our main
contributions are:
\begin{itemize}
  \item \textbf{Subject-level graph formulation:} a unified representation that
  aggregates all available vowels and syllables for each subject via a single
  kNN graph over SSL embeddings.
  \item \textbf{Comprehensive benchmarking:} a systematic evaluation of four SSL
  feature extractors and five GNN architectures on both Speech Analysis for Neurodegenerative Diseases (SAND) tasks under
  10-fold cross-validation (CV) and the official validation dataset.
  \item \textbf{Strong validation performance:} consistent gains over the SAND
  validation baselines by leveraging multi-recording fusion through graph-based
  message passing~\cite{sand_challenge_2025}.
\end{itemize}

\section{Methods}
\label{sec:method}

\subsection{Dataset}
\label{ssec:dataset}

The SAND Challenge dataset~\cite{sand_challenge_2025,verde2016vox4health,dubbioso2024voice} comprises 2{,}712
voice recordings from 339 Italian speakers, including 205 ALS patients and
134 healthy controls. Each subject provides five sustained vowel phonations
(\texttt{/a/, /e/, /i/, /o/, /u/}) and three DDK syllables with frequent
repetitions (\texttt{/pa/, /ta/, /ka/}). The official split includes 219
subjects for training and 53 for validation; the held-out test set is
evaluated via the challenge server and is not used in this study.

Following the challenge guidelines, we consider two tasks; Task~1: dysarthria
severity classification, formulated as a 5-class classification
problem, and Task~2: disease progression prediction , which predicts
the ALS Functional Rating Scale–Revised (ALSFRS-R) score~\cite{cedarbaum1999alsfrs} at the last visit using
recordings from initial visits (i.e., near diagnosis and several months
before the final assessment).

\subsection{Pre-processing}
\label{ssec:audio}

Each subject has up to eight recordings (five sustained vowels and three repetitions) originally sampled at 8\,kHz. To ensure
compatibility with pretrained SSL speech
encoders, we resample each recording to 16\,kHz. Waveforms are peak
normalized to the range $[-1, +1]$ and converted to a fixed-duration
20\,s clip by repeating (tiling) shorter recordings and truncating longer
ones. Each 20\,s waveform is then segmented into non-overlapping 2\,s
chunks, yielding exactly $T=10$ segments per recording. Consequently, a
subject with all eight recordings contributes up to $8 \times 10 = 80$
chunks in total.

\subsection{SSL Feature Extraction}
\label{ssec:ssl}

We used four transformer-based SSL encoders, all pre-trained on large
English speech corpora, namely the \emph{base} variants of wav2vec~2.0~\cite{baevski2020wav2vec2}, HuBERT~\cite{hsu2021hubert}, data2vec-audio~\cite{baevski2022data2vec}, and UniSpeech-SAT~\cite{chen2022unispeechsat}. 

Each 2-second chunk from the pre-processed recording, namely $\mathbf{s}_{i,j,t} \in \mathbb{R}^{1\times 32,000}$ for $i$-th participant's $j$-th utterance and $t$-th chunk, is passed through the frozen SSL encoder (last hidden
state, mean-pooled over time) to produce a 768-dimensional node feature
vector, namely $\mathbf{e}_{i,j,t} \in \mathbb{R}^{1\times768}$.

\subsection{Graph Construction}
\label{ssec:graph}
For each subject, we construct a single graph by aggregating all available
recordings. After SSL feature extraction, each
2\,s chunk is represented by a 768-dimensional embedding. Let
$\mathbf{e}_n \in \mathbb{R}^{768}$ denote the embedding of the $n$-th chunk
for a given subject, and stack all chunk embeddings for all utterances into a node-feature
matrix
$\mathbf{E} \in \mathbb{R}^{N \times 768}$, where $N = 10 \times R$ and
$R = 8$ is the number of available recordings (thus $N = 80$). We define the graph edges using a kNN graph in the embedding space based on cosine similarity. For two nodes with features
$\mathbf{e}_i$ and $\mathbf{e}_j$, we compute
\begin{equation}\label{eq:knn}
\begin{aligned}
  s_{ij} = \frac{\mathbf{e}_i^\top \mathbf{e}_j}{\lVert\mathbf{e}_i\rVert\,\lVert\mathbf{e}_j\rVert}, \quad
  \mathcal{E} = \bigl\{(i,j)\,:\, j \in \operatorname{kNN}_k(i)\bigr\},
\end{aligned}
\end{equation}
where $\operatorname{kNN}_k(i)$ gives the $k$ nodes with the highest
$s_{ij}$ for node $i$ (excluding $j=i$). We symmetrize the graph by adding
reciprocal edges and use $s_{ij}$ as the edge weight. This topology
captures acoustic proximity across phonation types and temporal segments,
allowing the GNN to propagate complementary information across chunks. In this study, to ensure the optimum solution is acquired, we run a hyperparameter tuning for $k \in \{1,3,5,10\}$ using grid search.

\begin{figure*}[!ht]
    \centering
    \includegraphics[width=0.8\linewidth,trim=0 6.5cm 0 4cm,clip]{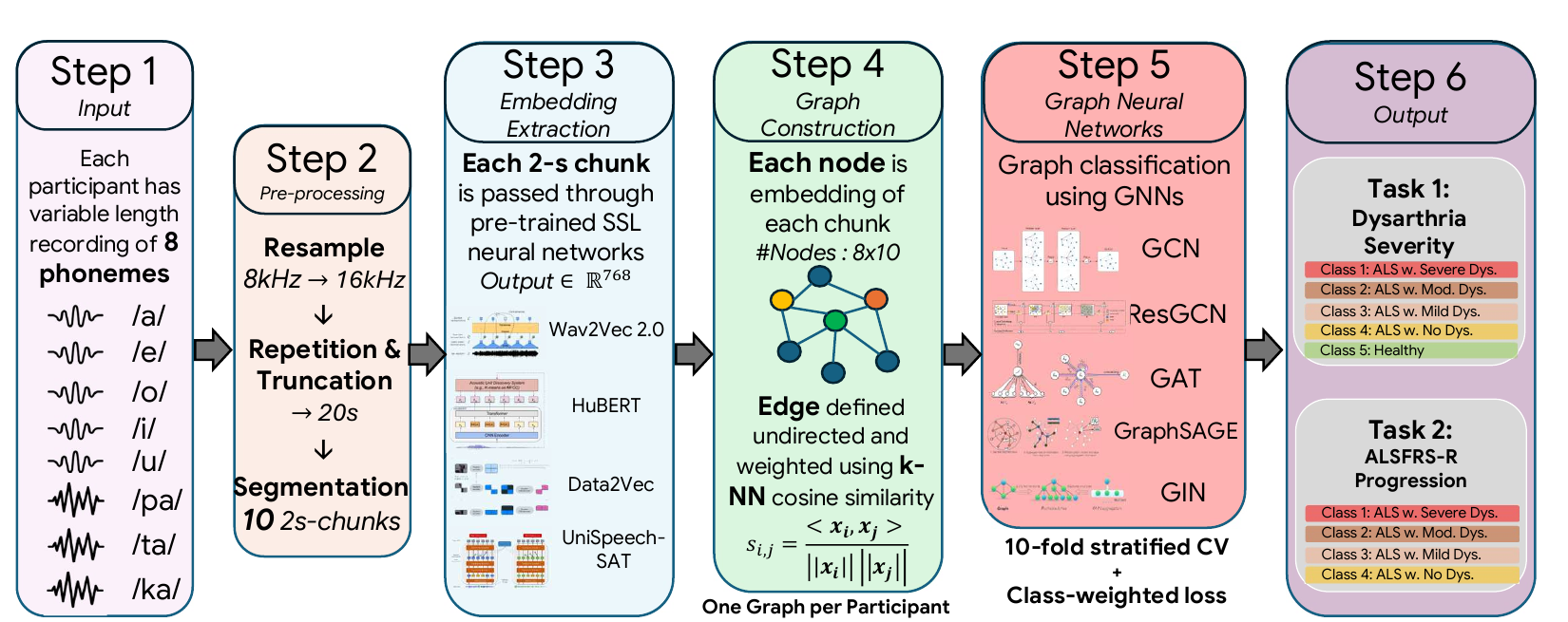}
  \caption{Proposed pipeline. For each patient, multiple recording
    segments are independently encoded by a frozen SSL model, assembled
    into a kNN graph, and classified via a GNN with global pooling.}
  \label{fig:arch}
\end{figure*}

\subsection{Graph Neural Networks}
\label{ssec:gnn}

We evaluate five GNNs under a unified
message-passing framework with a shared graph-level readout and classifier.
Let $\mathbf{H}^{(0)}=\mathbf{E}\in\mathbb{R}^{N\times d}$ denote the input
node features ($d=768$, $N= 80$). Each model
applies $L$ message-passing layers to produce node embeddings
$\mathbf{H}^{(L)}$. We obtain a graph representation by global mean pooling,
$\mathbf{z}=\frac{1}{N}\sum_{i=1}^{N}\mathbf{h}_i^{(L)}$, and feed $\mathbf{z}$
to a two-layer multilayer perceptron (MLP) with ReLU and dropout to predict class logits. After each
message-passing layer, we apply BatchNorm, ReLU, and dropout.

\begin{itemize}
  \item \textbf{GCN}~\cite{kipf2017gcn}: Graph Convolutional Network with
  renormalized first-order spectral filtering,
  \begin{equation}
    \mathbf{H}^{(l)} =
    \sigma\left(
      \tilde{\mathbf{D}}^{-\frac{1}{2}}
      \tilde{\mathbf{A}}
      \tilde{\mathbf{D}}^{-\frac{1}{2}}
      \mathbf{H}^{(l-1)} \mathbf{W}^{(l)}
    \right),
  \end{equation}
  where $\mathbf{A}\in\mathbb{R}^{N\times N}$ is the (weighted) adjacency matrix with $A_{ij}=s_{ij}$ for kNN-connected node pairs and $0$ otherwise, $\mathbf{H}^{(l)}$ denotes the node embeddings at layer $l$, $\tilde{\mathbf{A}}=\mathbf{A}+\mathbf{I}$ adds self-loops, and 
  $\tilde{D}_{ii}=\sum_j \tilde{A}_{ij}$ is the corresponding degree matrix. 
  \item \textbf{ResGCN}: A residual GCN variant with skip connections,
  \begin{equation}
    \mathbf{H}^{(l)} =
    \sigma\left(\operatorname{GCNConv}(\mathbf{H}^{(l-1)},\mathbf{A})\right)
    + \mathbf{H}^{(l-1)}.
  \end{equation}

  \item \textbf{GAT}~\cite{velickovic2018gat}: Graph Attention Network with
  multi-head attention. For each head $m$, node $i$ aggregates neighbors via
  attention weights $\alpha_{ij}^{(m)}$ and a learned projection
  $\mathbf{W}^{(m)}$:
  \begin{equation}
    \mathbf{h}_{i}^{(l,m)} =
    \sigma\left(
      \sum_{j\in\mathcal{N}(i)\cup\{i\}}
      \alpha_{ij}^{(m)} \mathbf{W}^{(m)} \mathbf{h}_{j}^{(l-1)}
    \right),
  \end{equation}
  and the final output is obtained by concatenating the $M$ head outputs,
  $\mathbf{h}_{i}^{(l)} = [\mathbf{h}_{i}^{(l,1)} \Vert \cdots \Vert \mathbf{h}_{i}^{(l,M)}]$.
  In this study, we used $M=4$ heads.

  \item \textbf{GraphSAGE}~\cite{hamilton2017sage}: Mean-aggregator GraphSAGE,
  \begin{equation}
    \mathbf{h}_i^{(l)} =
    \sigma\left(
      \mathbf{W}^{(l)}
      \left[\mathbf{h}_i^{(l-1)} \Vert
      \operatorname{mean}_{j\in\mathcal{N}(i)} \mathbf{h}_j^{(l-1)}\right]
    \right),
  \end{equation}
  which is well-suited to variable-sized graphs.

  \item \textbf{GIN}~\cite{xu2019gin}: Graph Isomorphism Network with sum
  aggregation and an MLP update,
  \begin{equation}
    \mathbf{h}_i^{(l)} =
    \operatorname{MLP}^{(l)}\left(
      (1+\epsilon)\mathbf{h}_i^{(l-1)} +
      \sum_{j\in\mathcal{N}(i)} \mathbf{h}_j^{(l-1)}
    \right),
  \end{equation}
  where $\epsilon$ is a learnable scalar.
\end{itemize}

To identify the best-performing configuration for each GNN, we performed a
grid search over the hidden dimension $d \in \{128,256\}$ and
the number of message-passing layers $L \in \{2,3\}$. Moreover, the dropout probability hyperparameter was selected using the grid-search $\in \{0.3,0.5\}$

\subsection{Training \& Benchmarking}
\label{ssec:training}

We first select model hyperparameters via stratified 10-fold CV over all available subjects. The best configuration is then retrained on
the official training split and benchmarked on the official validation split
provided by the SAND Challenge. The test set is evaluated exclusively through
the challenge server and is not publicly available; therefore, it is not used
in this study.

All models are trained with AdamW, cosine learning-rate annealing, and early
stopping (patience $=50$ epochs; maximum $200$ epochs). We use a batch size of
16 and weight decay of $10^{-4}$. The learning rate is selected by grid search from the subset of $\{10^{-3},3\times10^{-4}\}$. To address class imbalance in both
tasks, we apply inverse-frequency class weighting in the cross-entropy loss.
For 10-fold CV, we report macro F$_1$ (mF$_1$) as mean $\pm$ standard deviation across
folds. Following the recommendation of SAND Grand Challenge guidelines, for evaluation on the official validation split, we report balanced
accuracy, mF$_1$, and weighted F$_1$ (wF$_1$).

For both Task~1 (5-class severity) and Task~2 (4-class progression), we compute
per-class precision and recall from the confusion matrix and then accordingly per-class F$_1$:
\begin{equation}
\mathrm{P}_c = \frac{\mathrm{TP}_c}{\mathrm{TP}_c+\mathrm{FP}_c}, 
\mathrm{R}_c = \frac{\mathrm{TP}_c}{\mathrm{TP}_c+\mathrm{FN}_c},
\mathrm{F_1}_c = \frac{2\,\mathrm{P}_c\,\mathrm{R}_c}{\mathrm{P}_c+\mathrm{R}_c}.
\end{equation}
mF$_1$ averages equally across classes:
\begin{equation}
\mathrm{mF_1} = \frac{1}{C}\sum_{c=1}^{C}\mathrm{F1}_c,
\end{equation}
where $C=5$ for Task~1 and $C=4$ for Task~2. wF$_1$ weights
each class by its support $n_c$, which denotes the support of class $c$ (the number of samples with true
label $c$ in the evaluation split), and $N=\sum_{c=1}^{C} n_c$ is the total
number of evaluated samples:
\begin{equation}
\mathrm{wF_1} = \sum_{c=1}^{C}\frac{n_c}{\sum_{c'=1}^{C} n_{c'}}\,\mathrm{F_1}_c.
\end{equation}
Balanced accuracy (BACC) is the mean per-class recall:
\begin{equation}
\mathrm{BACC} = \frac{1}{C}\sum_{c=1}^{C}\mathrm{R}_c.
\end{equation}

All experiments are conducted on a single NVIDIA RTX 5080 GPU with 16\,GB VRAM.
Models are implemented in Python~3.11.9 using PyTorch~2.10.0 (CUDA~13.0) and
HuggingFace Transformers~5.1.0.
 The overall proposed pipeline is shown in Fig. \ref{fig:arch}.

\section{Results}
\label{sec:results}

\subsection{Task~1: Dysarthria Severity Classification}
\label{ssec:task1}

\begin{figure}[!ht]
    \centering
    \subfigure{
        \includegraphics[width=\columnwidth, trim=0 2.5cm 0 0, clip]{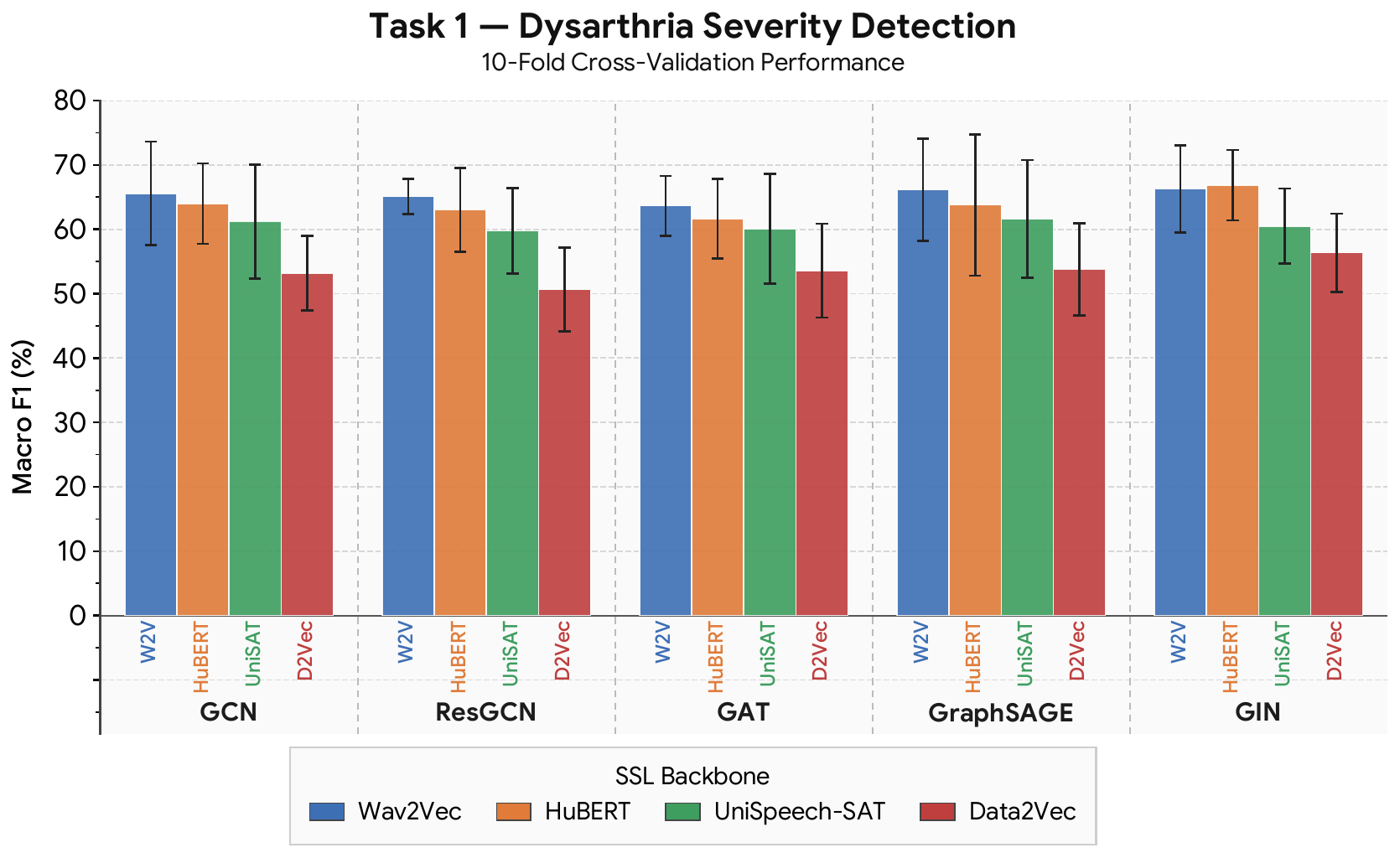}
    }
    
    \subfigure{
        \includegraphics[width=\columnwidth]{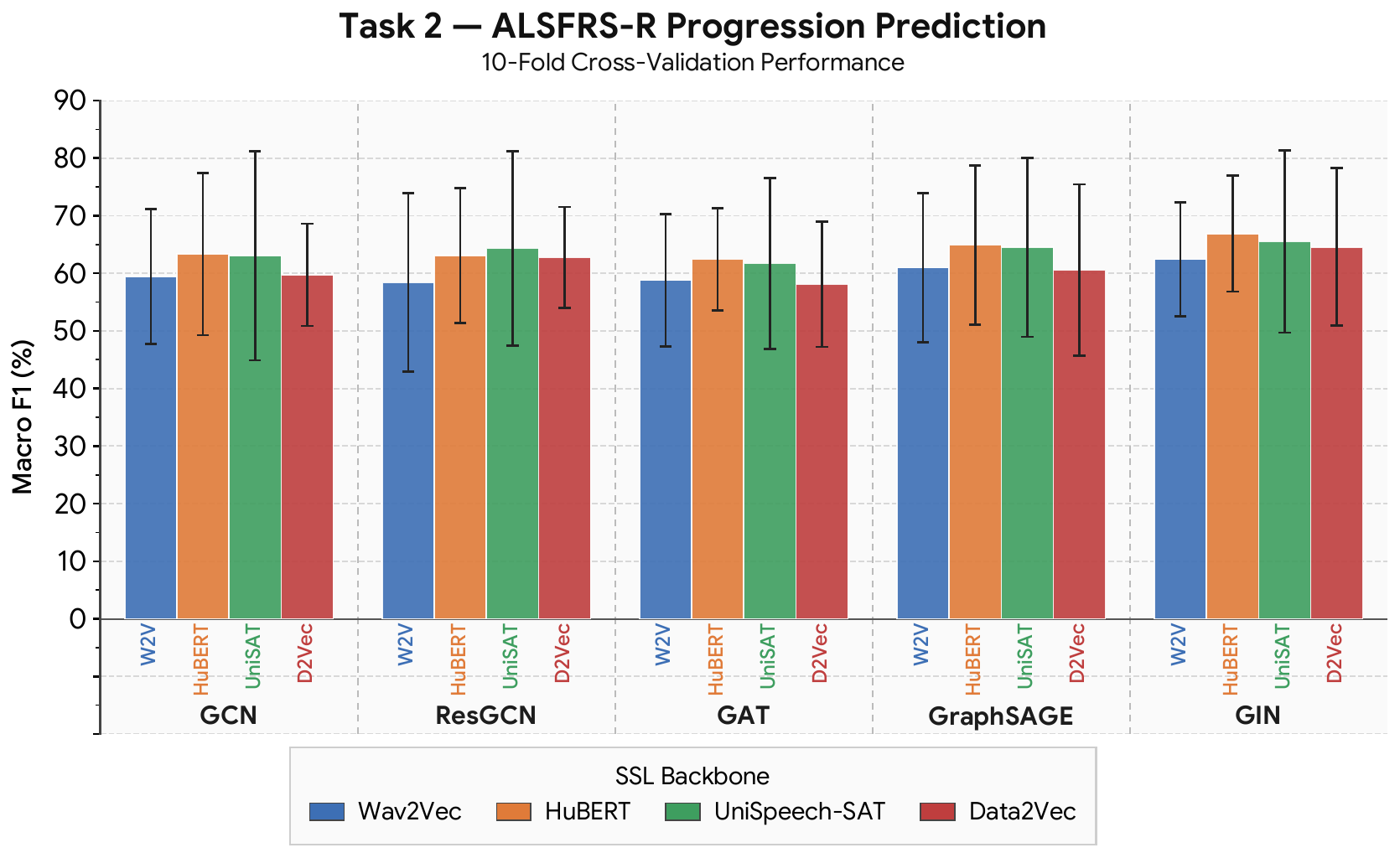}
    }

    \vspace{-4mm}
    \caption{Best 10-fold cross-validation mF$_1$ for both tasks per SSL $\times$ GNN combination (error bars denote CV $\sigma$).}
    \label{fig:task1_task2_macroF1}
\end{figure}

Although Task~1 involves a larger label space (5 classes), it is generally
easier than Task~2 because it relies on contemporaneous acoustic cues of
dysarthria rather than forecasting future functional decline,
acoustic markers of which can be weaker and confounded by
inter-subject variability and heterogeneous progression trajectories. The results in
Table~\ref{tab:main} and Fig.~\ref{fig:task1_task2_macroF1} support this
expectation, with consistently higher scores for Task~1 across models.

Across both the 10-fold CV protocol and the official validation split, the
best-performing configuration is GIN paired with HuBERT
embeddings. It achieves a mF$_1$ of $0.67 \pm 0.05$ across CV folds, and
on the official validation split attains mF$_1$\,=\,0.73, wF$_1$\,=\,0.67, and
BACC\,=\,0.72. The second-best model is GIN with
wav2vec~2.0, yielding $0.66 \pm 0.07$ mF$_1$ in CV and mF$_1$\,=\,0.71
on the validation split.

Overall, GIN provides the strongest and most consistent performance across SSL
front-ends for Task~1, suggesting that its sum-aggregation and MLP update are
well suited to capturing subject-specific phonetic variability across vowels
and diadochokinetic syllables. Across GNN backbones, HuBERT embeddings
generally perform best, which we hypothesize is due to HuBERT's masked
prediction objective that encourages learning phoneme-discriminative and
speaker-robust representations, benefiting severity classification under
limited labeled data.

\begin{table*}[!ht]
\centering
\caption{Best performed model on mF$_1$ score on the validation set. Each configuration per SSL $\times$ GNN for Task 1 and Task 2 is selected by highest validation mF$_1$. Gray cells indicate the best SSL embedding per GNN (within each task). Orange cells indicate the best overall (GNN, SSL) per task based on mF$_1$.}
\label{tab:main}
\resizebox{\textwidth}{!}{%
\begin{tabular}{c| c| ccccc| ccccc| ccccc| ccccc| ccccc}
\hline
\multirow{2}{*}{\textbf{Task}} & \multirow{2}{*}{\textbf{SSL}}
& \multicolumn{5}{c|}{\textbf{GCN}}
& \multicolumn{5}{c|}{\textbf{ResGCN}}
& \multicolumn{5}{c|}{\textbf{GAT}}
& \multicolumn{5}{c|}{\textbf{GraphSAGE}}
& \multicolumn{5}{c}{\textbf{GIN}} \\
&
& \textbf{\boldmath$k$} & \textbf{\boldmath$d$} & \textbf{mF$_1$} & \textbf{wF$_1$} & \textbf{BACC}
& \textbf{\boldmath$k$} & \textbf{\boldmath$d$} & \textbf{mF$_1$} & \textbf{wF$_1$} & \textbf{BACC}
& \textbf{\boldmath$k$} & \textbf{\boldmath$d$} & \textbf{mF$_1$} & \textbf{wF$_1$} & \textbf{BACC}
& \textbf{\boldmath$k$} & \textbf{\boldmath$d$} & \textbf{mF$_1$} & \textbf{wF$_1$} & \textbf{BACC}
& \textbf{\boldmath$k$} & \textbf{\boldmath$d$} & \textbf{mF$_1$} & \textbf{wF$_1$} & \textbf{BACC} \\
\hline
\multirow{4}{*}{1} & \textit{Wav2Vec 2.0}
& 5  & 128 & 0.66 & 0.62 & 0.69
& 10 & 256 & 0.68 & 0.65 & 0.72
& 10 & 256 & 0.64 & 0.59 & 0.63
& 1  & 256 & 0.66 & 0.62 & 0.66
& 10 & 128 & 0.71 & 0.62 & 0.69 \\
& \textit{HuBERT}
& \bestcell{1}  & \bestcell{128} & \bestcell{0.71} & \bestcell{0.66} & \bestcell{0.72}
& \bestcell{5}  & \bestcell{128} & \bestcell{0.71} & \bestcell{0.62} & \bestcell{0.69}
& \bestcell{10} & \bestcell{256} & \bestcell{0.67} & \bestcell{0.66} & \bestcell{0.68}
& \bestcell{5}  & \bestcell{128} & \bestcell{0.72} & \bestcell{0.66} & \bestcell{0.72}
& \bestoverall{1}  & \bestoverall{128} & \bestoverall{0.73} & \bestoverall{0.67} & \bestoverall{0.72} \\
& \textit{UniSpeech-SAT}
& 10 & 256 & 0.61 & 0.59 & 0.63
& 5  & 256 & 0.54 & 0.58 & 0.59
& 3  & 128 & 0.52 & 0.56 & 0.57
& 3  & 128 & 0.62 & 0.64 & 0.66
& 10 & 128 & 0.62 & 0.58 & 0.62 \\
& \textit{Data2Vec}
& 1  & 128 & 0.60 & 0.62 & 0.58
& 3  & 128 & 0.60 & 0.65 & 0.62
& 1  & 256 & 0.60 & 0.66 & 0.63
& 3  & 128 & 0.57 & 0.62 & 0.54
& 3  & 128 & 0.60 & 0.60 & 0.59 \\
\hline
\multirow{4}{*}{2} & \textit{Wav2Vec 2.0}
& \bestcell{3}  & \bestcell{256} & \bestcell{0.64} & \bestcell{0.65} & \bestcell{0.67}
& \bestcell{10} & \bestcell{256} & \bestcell{0.64} & \bestcell{0.62} & \bestcell{0.61}
& \bestcell{10} & \bestcell{128} & \bestcell{0.62} & \bestcell{0.57} & \bestcell{0.67}
& \bestcell{10} & \bestcell{256} & \bestcell{0.64} & \bestcell{0.61} & \bestcell{0.64}
& 10 & 256 & 0.64 & 0.65 & 0.64 \\
& \textit{HuBERT}
& 5  & 256 & 0.63 & 0.62 & 0.64
& 3  & 128 & 0.59 & 0.56 & 0.57
& 10 & 256 & 0.56 & 0.57 & 0.61
& 5  & 128 & 0.63 & 0.58 & 0.62
& \bestoverall{10} & \bestoverall{256} & \bestoverall{0.69} & \bestoverall{0.68} & \bestoverall{0.65} \\
& \textit{UniSpeech-SAT}
& 1  & 256 & 0.59 & 0.59 & 0.62
& 1  & 128 & 0.60 & 0.60 & 0.62
& 1  & 128 & 0.54 & 0.53 & 0.57
& 3  & 256 & 0.61 & 0.62 & 0.58
& 3  & 256 & 0.68 & 0.64 & 0.72 \\
& \textit{Data2Vec}
& 5  & 128 & 0.49 & 0.50 & 0.49
& 3  & 128 & 0.54 & 0.53 & 0.51
& 1  & 128 & 0.54 & 0.46 & 0.54
& 1  & 128 & 0.49 & 0.49 & 0.55
& 10 & 256 & 0.53 & 0.57 & 0.56 \\
\hline
\end{tabular}%
}
\end{table*}

\subsection{Task~2: ALSFRS-R Progression Prediction}
\label{ssec:task2}

Consistent with Task~1 (Table~\ref{tab:main}, Fig.~\ref{fig:task1_task2_macroF1}), the best-performing configuration in both 10-fold CV
and the official validation split is GIN paired with HuBERT
embeddings. It achieves a mF$_1$ of $0.67 \pm 0.10$ across CV folds and,
on the validation split, attains mF$_1$\,=\,0.69, wF$_1$\,=\,0.68, and
BACC\,=\,0.65. The second-best model is GIN with
UniSpeech-SAT, yielding $0.65 \pm 0.16$ mF$_1$ in CV and
mF$_1$\,=\,0.68 on the validation split.

Overall, GIN remains the most effective GNN backbone for Task~2,
suggesting that sum-based aggregation is beneficial when informative cues are
distributed across a subset of chunks and must be combined at the subject
level. Among the remaining GNN architectures (GCN, ResGCN, GAT, and GraphSAGE),
wav2vec~2.0 tends to provide the strongest mF$_1$, indicating that its
representations transfer well to progression prediction when paired with less
expressive graph aggregators.

\subsection{Comparison with SAND Challenge Baselines}

Our best configuration on the validation set yields 0.73 mF$_1$ for
Task~1 (HuBERT + GIN) and 0.69 for Task~2
(HuBERT + GIN), compared to the SAND leaderboard's top scores
of 0.61 and 0.58 on the held-out test set.  We caution that a direct
comparison is confounded by the partition difference, our numbers are
computed on the internal validation split rather than the final test set.
Nonetheless, the comparison with the baseline of mF$_1$ of 0.61 for Task 1 and 0.58 for Task 2 on the same validation set \cite{sand_challenge_2025} and  consistent margin across all our configurations suggests
that the graph-based multi-segment modelling provides strong representation that
conventional per-recording approaches do not exploit.

\begin{table}[!ht]
  \centering
  \caption{Comparison with SAND challenge baselines and leaderboard.
           Best result per column in \textbf{bold}.}
  \label{tab:comparison}
  \resizebox{\linewidth}{!}{%
  \begin{tabular}{lccc}
    \toprule
    \textbf{Method} & \textbf{Task~1 mF$_1$} & \textbf{Task~2 mF$_1$} & \textbf{Set} \\
    \midrule
    SAND Baseline (ViT / PART)~\cite{sand_challenge_2025}
                           & 0.61 & 0.58 & Val \\
    TUKE~(1st, Task~1)~\cite{sand_challenge_2025}
                           & 0.61& --              & Test \\
    ISDS~(1st, Task~2)~\cite{sand_challenge_2025}
                           & 0.51              & 0.58 & Test \\
    \midrule
    Ours (HuBERT + GIN)    & \textbf{0.73} & $\mathbf{0.69}$ & Val \\
    Ours (Wav2Vec 2.0 + GIN)    & 0.71 & $0.64$ & Val \\
    Ours (UniSpeech-SAT + GIN)    & 0.62 & $0.68$ & Val \\
    Ours (Data2Vec + GAT)    & 0.60 & $0.53$ & Val \\

    \bottomrule
  \end{tabular}
  }
\end{table}

\section{Discussion}
\label{sec:discussion}
We proposed a subject-level graph learning pipeline that converts multiple short speech segments per speaker into a single kNN graph in SSL-embedding space and performs graph classification using standard GNN backbones. Across both SAND tasks, the same pairing, HuBERT embeddings with a GIN classifier, consistently delivered the strongest validation mF$_1$ (0.73 for Task~1; 0.69 for Task~2), outperforming the SAND baselines reported for the official validation split (ViT for Task~1 and PART for Task~2)~\cite{sand_challenge_2025}. These results suggest that explicitly aggregating complementary evidence across heterogeneous recordings (vowels and DDK syllables) and across multiple temporal chunks provides a more informative subject representation than treating recordings independently.

Two design choices appear to be particularly effective: (i) HuBERT features and (ii) sum-based graph aggregation. HuBERT learns representations by masked prediction of hidden units, which encourages phonetic structure discovery and robust intermediate units even with limited downstream labels~\cite{hsu2021hubert}. This property is plausible to help in ALS dysarthria tasks where discriminative cues may be subtle and distributed across phonation types. On the graph side, GIN is among the most expressive message-passing architectures, with a sum aggregator and MLP updates that can preserve multiset information from neighborhoods better than mean-based alternatives in graph classification settings~\cite{xu2019gin}. In our setting, informative chunks may be sparse (e.g., only a subset of vowels or brief temporal regions show clear impairment), sum aggregation can amplify these local cues when pooling to a subject-level decision, which may explain GIN's advantage over GCN/GAT/GraphSAGE across SSL front-ends. Previously, Sheikh \textit{et al.} investigated the potential of graph neural networks in detecting Parkinson's disease and showed that by representing speech segments as nodes and capturing the similarity between segments through edges, the GCN model facilitates the aggregation of dysarthric cues across the graph, effectively exploiting segment relationships and mitigating the impact of label noise \cite{sheikh2025graph}.

Task~1 (severity at time of recording) is consistently easier than Task~2 (predicting ALSFRS-R outcome at the final visit from early recordings), aligning with the clinical intuition that current dysarthria is more directly observable in acoustics than future functional decline, which is confounded by heterogeneous progression trajectories and non-speech factors~\cite{cedarbaum1999alsfrs}. The performance gap between tasks also indicates that the SAND progression labels encode a more challenging mapping from speech to longitudinal disease evolution. Nevertheless, the fact that the same graph-based approach remains competitive in Task~2 supports the hypothesis that multi-recording fusion is beneficial for progression modeling as well, likely because it stabilizes predictions against within-subject variability and recording-specific noise.

Our graphs are constructed by connecting chunk embeddings that are close under cosine similarity, rather than by a predefined temporal chain or by grouping solely within the same utterance type. This design encourages cross-recording information flow (e.g., between \texttt{/a/} and \texttt{/i/} chunks) and may better capture subject-specific acoustic manifolds spanning different tasks. Prior clinical work suggests that sustained vowels can be particularly sensitive to ALS-related motor speech changes~\cite{tomik1999acoustic}, and the Vox4Health-style protocol used to collect these recordings was designed to enable scalable voice monitoring~\cite{verde2016vox4health,dubbioso2024voice}. In that context, allowing message passing across all phonations may help the model combine complementary cues (source, filter, and stability/irregularity markers) that manifest differently across vowels and DDK.

On the official validation split, our best models exceed the baseline macro or averaged F$_1$ reported by the challenge organizers for both tasks~\cite{sand_challenge_2025}. While a direct comparison to leaderboard test performance is not appropriate due to split differences and server-only evaluation, the consistent margin over the validation baseline across many combinations suggests that the improvement is not attributable to a single lucky configuration. Instead, it points to a more general advantage of subject-level multi-segment modeling over per-recording pipelines, especially when the dataset contains multiple elicitation tasks per speaker.

There are several limitations for this study that should be acknowledged. First, all SSL encoders are pretrained primarily on large English corpora, whereas SAND recordings are Italian; cross-lingual mismatch may limit absolute performance and may partly explain why different SSL models behave inconsistently across GNN backbones~\cite{baevski2020wav2vec2,hsu2021hubert,baevski2022data2vec,chen2022unispeechsat}. Second, we used simple tiling/truncation to reach a fixed 20\,s duration, which can introduce artificial repetition and potentially bias neighborhood construction by creating highly similar segments. Third, our kNN topology is fixed and unsupervised; it may connect segments by nuisance similarity (e.g., channel characteristics) rather than clinically meaningful proximity. Finally, our evaluation focuses on the official validation split, and test-set generalization must be confirmed through challenge server submission.


\section{Conclusion}
\label{sec:conclusion}

We introduced a simple but effective way to turn multiple short phonation
recordings from a single speaker into a unified, subject-level representation;
a kNN graph built in the space of frozen SSL embeddings and
classified with a graph neural network. Across an extensive sweep of
different configurations per task on the SAND dataset, the same design choice
repeatedly emerged as the most reliable, namely pairing HuBERT embeddings with a GIN
backbone. This combination achieved the strongest validation performance for
both dysarthria severity classification and ALSFRS-R progression prediction,
substantially improving over the challenge baselines and highlighting the value
of graph-based fusion across recording types and temporal segments. Overall,
these results suggest that subject-level aggregation in embedding space graph representation is a
promising direction for scalable ALS speech assessment, especially in
low-resource settings where labeled clinical data remain limited.

\section{Generative AI Use Disclosure}
The authors take full responsibility for the accuracy, originality, and integrity of the final work and affirm that generative AI was used solely for language refinement and drafting assistance.

\bibliographystyle{IEEEtran}
\bibliography{ref}

\end{document}